\documentclass[hidelinks,conference]{IEEEtran}
\IEEEoverridecommandlockouts
\usepackage{amsmath,amssymb,amsfonts}
\usepackage{algorithmic}
\usepackage{graphicx}
\usepackage{textcomp}
\usepackage{xcolor}

\usepackage[backend=biber,style=ieee,natbib=true]{biblatex} 
\usepackage{hyperref}

\usepackage{booktabs}
\usepackage{tabularx}

\usepackage[english]{babel}   % load english localization
\usepackage[T1]{fontenc}      % correct some font representation,

\graphicspath{{figs/}}

\usepackage[most]{tcolorbox}
\newtcolorbox{myboxii}[1][]{
	breakable,
	freelance,
	title=#1,
	colback=white,
	colbacktitle=white,
	coltitle=black,
	fonttitle=\bfseries,
	bottomrule=0pt,
	boxrule=0pt,
	colframe=white,
	overlay unbroken and first={
		\draw[black!75!black,line width=1pt]
		([xshift=5pt]frame.north west) -- 
		(frame.north west) -- 
		(frame.south west);
		\draw[black!75!black,line width=1pt]
		([xshift=-5pt]frame.north east) -- 
		(frame.north east) -- 
		(frame.south east);
	},
	overlay unbroken app={
		\draw[black!75!black,line width=1pt,line cap=rect]
		(frame.south west) -- 
		([xshift=5pt]frame.south west);
		\draw[black!75!black,line width=1pt,line cap=rect]
		(frame.south east) -- 
		([xshift=-5pt]frame.south east);
	},
	overlay middle and last={
		\draw[black!75!black,line width=1pt]
		(frame.north west) -- 
		(frame.south west);
		\draw[black!75!black,
		line width=1pt]
		(frame.north east) -- 
		(frame.south east);
	},
	overlay last app={
		\draw[black!75!black,line width=1pt,line cap=rect]
		(frame.south west) --
		([xshift=5pt]frame.south west);
		\draw[black!75!black,line width=1pt,line cap=rect]
		(frame.south east) --
		([xshift=-5pt]frame.south east);
	},
}

\newcommand{\pageClimb}{\url{https://github.com/gamedev-studies/jungle-climb}}

\def\BibTeX{{\rm B\kern-.05em{\sc i\kern-.025em b}\kern-.08em
    T\kern-.1667em\lower.7ex\hbox{E}\kern-.125emX}}

\begin{document}

\title{Assessing Video Game Balance using Autonomous Agents}

\author
{
\IEEEauthorblockN{1\textsuperscript{st} Cristiano Politowski}
\IEEEauthorblockA{\textit{École de Technologie Supérieure} \\
Montreal, Quebec, Canada \\
cristiano.politowski@etsmtl.ca}
\and
\IEEEauthorblockN{2\textsuperscript{nd} Fabio Petrillo}
\IEEEauthorblockA{\textit{École de Technologie Supérieure} \\
Montreal, Quebec, Canada \\
fabio.petrillo@etsmtl.ca}
\and
\IEEEauthorblockN{3\textsuperscript{rd} Ghizlane ElBoussaidi}
\IEEEauthorblockA{\textit{École de Technologie Supérieure} \\
Montreal, Quebec, Canada \\
ghizlane.elboussaidi@etsmtl.ca}\\
\and
\IEEEauthorblockN{4\textsuperscript{th} Gabriel C. Ullmann}
\IEEEauthorblockA{
\textit{Concordia University} \\
Montreal, Quebec, Canada \\
g\_cavalh@live.concordia.ca}\\
\and
\IEEEauthorblockN{5\textsuperscript{th} Yann-Ga\"el Gu\'{e}h\'{e}neuc}
\IEEEauthorblockA{\textit{Concordia University} \\
Montreal, Quebec, Canada \\
yann-gael.gueheneuc@concordia.ca}
}

\maketitle

\begin{abstract}
As the complexity and scope of games increase, game testing, also called playtesting, becomes an essential activity to ensure the quality of video games. Yet, the manual, ad-hoc nature of game testing leaves space for automation. In this paper, we research, design, and implement an approach to supplement game testing to balance video games with autonomous agents. We evaluate our approach with two platform games. We bring a systematic way to assess if a game is balanced by (1) comparing the difficulty levels between game versions and issues with the game design, and (2) the game demands for skill or luck.
\end{abstract}


\section{Introduction}

% context
Game development is an iterative process \cite{politowskiAreOldDays2016}. Game developers start with a core mechanic, limited in scope, and then iterate, adding new features until they deem the game complete. For every new change in the game, game testers (playtesters) interact with the game and provide feedback to the game developers who then change the game. They use experimentation and trial-and-error to tweak the game mechanics and make it engaging for the players. Keeping the game challenging while avoiding boredom requires \textit{balancing} it \cite{schell_art_2008}, which is hard to translate to the actual game specification and relies on empirical knowledge. This constant experimentation implies there are no clear requirements but a ``vision'' for games.

In this paper, we propose an approach to assess video game balance (semi) automatically. Instead of manually testing games, we propose an automated approach with autonomous agents to aid game developers assess the game's balance. \citet{schell_art_2008} lists 12 common types of balance in video games. In this paper, we focus on two balance types: \textit{Challenge vs.\ Success} about keeping the player engaged considering the game difficulty and the player's skills and \textit{Skill vs.\ Chance} about needing luck instead of skills to succeed. Games of skills are more like athletic contests (which player is the best?) while games of chance are more random. For example, dealing out a hand of cards is a pure chance but choosing how to play them is pure skill.

Similar to unit tests in continuous integration pipelines in traditional software development, with which a system warns developers when a test fails, we want to provide developers with an automated process that would warn them when a new version of their game is too far from a given balance. Our approach brings a systematic and autonomous way to identify if the game is balanced by (1) checking the difficulty spikes and (2) if the game demands more skill or luck.

% method
% We want the game development to become an iterative process, similar to the use of Continuous Integration (CI) pipelines. For instance, the developer writes a unit test for each new feature. By adding these tests to the CI system the developer automates its execution. Thus, for each new feature, the CI system warns the developer if there is any failure with the previous tests. Instead, in our case, we write integration tests that warn the game developer if there is an issue with the game balance. To do so, we use autonomous agents to aid the process of testing to balance the game mechanics. 

% why DRL
To do so, we train autonomous agents using Deep Reinforcement Learning (DRL). Training agents to play games with DRL is an ever-increasing research area \cite{mnih2013playing} and machine learning models allow autonomous agents to master games. Video games offer an environment with reduced scope (compared to real life) that suits the training of autonomous agents. These approaches show great success in mastering simple and complex games \cite{Justesen2020}.
% \footnote{\url{https://openai.com/projects/five/}} games. %We use autonomous agents because (1) scripts are immutable and keeping them up to date with the game modifications requires much effort; and (2) agents play differently every time, which helps to explore different game states.
% \Ghiz{Is it possible to add a sentence or two to stress the novelty of the approach compared to existing related work?}
In our approach, we incorporate the agents into the game development process and provide a solution for testing the game balance. Most game studios, especially the small ones, do not have the time or the budget to adopt complex and costly solutions. Yet, they can benefit from a more feasible approach.

% Because games are big in scope, we use a scenario representing a chunk of the game. This scenario has different game attributes. In unit testing, the oracle is well defined and immutable. For balancing the game we need to be less strict because the balance of the game depends on the game developer's vision, which is subjective. Therefore, we use metrics from within the game to assert its balance. For example, we verify if the score is too high or low and if the random agent is performing better than the others.

% With this automated process, developers can quickly test different attributes of their game automatically and in parallel. It provides evidence to developers on which changes benefit their game. It thus saves time, reduces manual effort, reduces the cost, and also helps improve the game's quality.

In the following, we describe our approach, how we implemented it, and how we validated our testing approach with two platform games. 
Section~\ref{sec:related-works} discusses the related works and how our approach differs from them.
Section~\ref{sec:approach} presents the approach.
Section~\ref{sec:implementation} shows how we technically implemented the approach. 
Section~\ref{sec:case-batkill} and Section~\ref{sec:case-jungle} are the case studies that use our approach.
Section~\ref{sec:discussion} presents the discussion and Section~\ref{sec:threats} the Threats to Validity.
Finally, Section~\ref{sec:conclusion} shows the conclusion and future works.

%%%
% RELATED WORK
%%%

\section{Related Work} \label{sec:related-works}

%There is a large community building DRL models to master games (play as well as a human) \cite{Justesen2020}, as for example, DeepMind's Agent57 model\footnote{\url{https://deepmind.com/blog/article/Agent57-Outperforming-the-human-Atari-benchmark}}. Yet, we saw that, for game testing purpose, simple models like Proximal Policy Optimization (PPO) \cite{Schulman17} and Asynchronous Advantage Actor-Critic (A2C) \cite{Mnih2016} are commonly used.
% PPO\footnote{\textit{Proximal Policy Optimization (PPO)} is a Reinforcement Learning algorithm ``which alternates between sampling data through interaction with the environment and optimizing a `surrogate' objective function using stochastic gradient ascent'' \cite{Schulman17}. The algorithm aims to maximize the probability of a set of actions being taken by the agent, given these actions make the agent get rewards above average during its interaction with the environment. PPO is an on-policy algorithm, meaning it learns by comparing the current set of actions taken with the previous one, without using a replay memory. } and 
% A2C\footnote{\textit{A2C} is a synchronous variant of Asynchronous Advantage Actor-Critic (A3C) that uses agents running in parallel to explore different parts of the environment \cite{Mnih2016}. By doing so, the algorithm does not need to use a replay memory. Similar to PPO, after reaching a terminal state (game over) or a maximum number of actions, the algorithm updates its policy, which is the function that generates the set of actions to be taken by the agent. This update is done to make the policy more likely to generate actions that will lead to high rewards.} are commonly used.

We read 199 papers about video game testing\footnote{\url{https://doi.org/10.5281/zenodo.7768876}} but, for the sake of space, here we will show the ones more close to the our testing objective: video game balancing. 
%Most of the papers lack a complete solution for video game testing. They propose theoretical approaches focused on the modelling and training of autonomous agents, providing few details about the assessment of the game. The great majority of the papers did not provided the source code. 
In the following we summarize, to the best of our knowledge, all the papers that relate to video game balancing.

\citet{isaksenExploringGameSpace2018} used A-Star, MCTS and visualization to determine the game difficulty by
verifing different attributes of the same game without changing its rules. The authors use metrics to predict the score of a platformer called Flappy bird.
\citet{gudmundssonHumanLikePlaytestingDeep2018}	tested the difficulty of a game level using autonomous agents, powered by a CNN, to simulate gameplay and the ``success rate'' against human players. They validated using a Match-3 game. They wanted to predict the difficulty of a new game level automatically. They claimed that the difficulty of new levels could be tested automatically. 
\citet{roohiPredictingGameDifficulty2020} predicted the pass rate (win the game) and churn (abandon the game) in new levels of a Match-3 game. They used gameplay data from autonomous agents and playtesters. The agents were trained using PPO within the Unity-ML engine. They claimed that the pass rate and churn of new levels could be tested automatically.

\citet{delaurentisAutomatedGameBalance}	defined a framework that predicted the game balance using data from autonomous agents, trained with CNN and MCTS, to play against each other in a RTS multiplayer game. They assessed the playsession with a visualization tool together with the actions performed by the agents.
\citet{pfauDungeonsReplicantsAutomated2020, Pfau2022Dungeons2} used Deep Player Behavior Modeling (DPBM) and data from real players to model autonomous agents and play a MMO game. The idea was to replicate the human behaviour and then assess the playsession manually.
\citet{demesentiersilvaAIbasedPlaytestingContemporary2017, silvaAIEvaluatorSearch} created agents using A-Star and MCTS to explore a board game called Ticket to Ride. The focus was more exploring than dealing with the game balance.

% \citet{mugraiAutomatedPlaytestingMatching2019} proposed an agent based on GA and MCT to create different procedural personas and human play styles for a Match-3 game. They focused on modeling the agents rather than the testing part.
% %
% \citet{johnAdversarialBehaviourDebugging} used NEAT and the Unity engine to create autonomous agents to play three different games. The authors then used the record-and-replay technique to assess the playsession. 
% \citet{zhaoLightweightApproachHumanLike} also used record-and-replay to balance a mobile game.
% %
% \citet{garca-snchezAutomatedPlaytestingCollectible2018} used Evolutionary algorithm to create recommendations of Hearthstone decks.
% %
% \citet{shinPlaytestingMatchGame2020} used MCTS and CNN to model agents to play a Match-3 game. The goal is to provide a easy way so the designer can assess, manually, the level difficulty.

%\citet{zhaoWinningIsnEverything2020} goals are to use autonomous agents to provide feedback during the game's development...

\citet{liuAutomaticGenerationTower2019} used procedural content generation to create new tower defense game levels to then playtest it autonomously using flat MCTS.
\citet{morosanAutomatedGameBalancing2017,  morosanLessonsTestingEvolutionary2018} used evolutionary algorithm to balance different games, like Pacman and StarCraft, using the win-rate as oracle.
\citet{preussIntegratedBalancingRTS2018} used the feature of the open-source game OpenRA, an RTS, to balance the strategies the player can adopt.

%\citet{birkMetricAutomaticallyFlagging2015} did not used autonomous agents to play the game but created a metric called Death-Related Problem Likelihood Indicator (DPLI) to spot the frequency of death in levels of the game. The authors then used logs and heatmaps to manually assess the issues.
%%
%\citet{paduraruAutomatedGameTesting} automatically create behaviors for game agents of different difficulty classes. Then used behaviour trees to automatic test the game difficulty.
%%
%\citet{zookMonteCarloTreeSearch2019} used active learning, that is, selecting among a set of possible inputs to get the best output, for parameters tunning the game.

% what i have different
%\begin{myboxii}[]
%The previous works create autonomous agents to master games but did not incorporate the agents into the game development process nor provide approaches for testing balance or difficulty. %Most game studios, especially the small ones, do not have the time or the budget to adopt complex and costly solutions like these, but they can benefit from a more feasible approach. Therefore, instead of thinking of game testing as an isolated process, we aim to the process of finding the right balance of the game.
% \cris{Make the gaps explicit criteria.}
%\end{myboxii}

\section{Approach} \label{sec:approach}

\begin{figure*}[!htb]
	\centering
	\includegraphics[width=.8\textwidth]{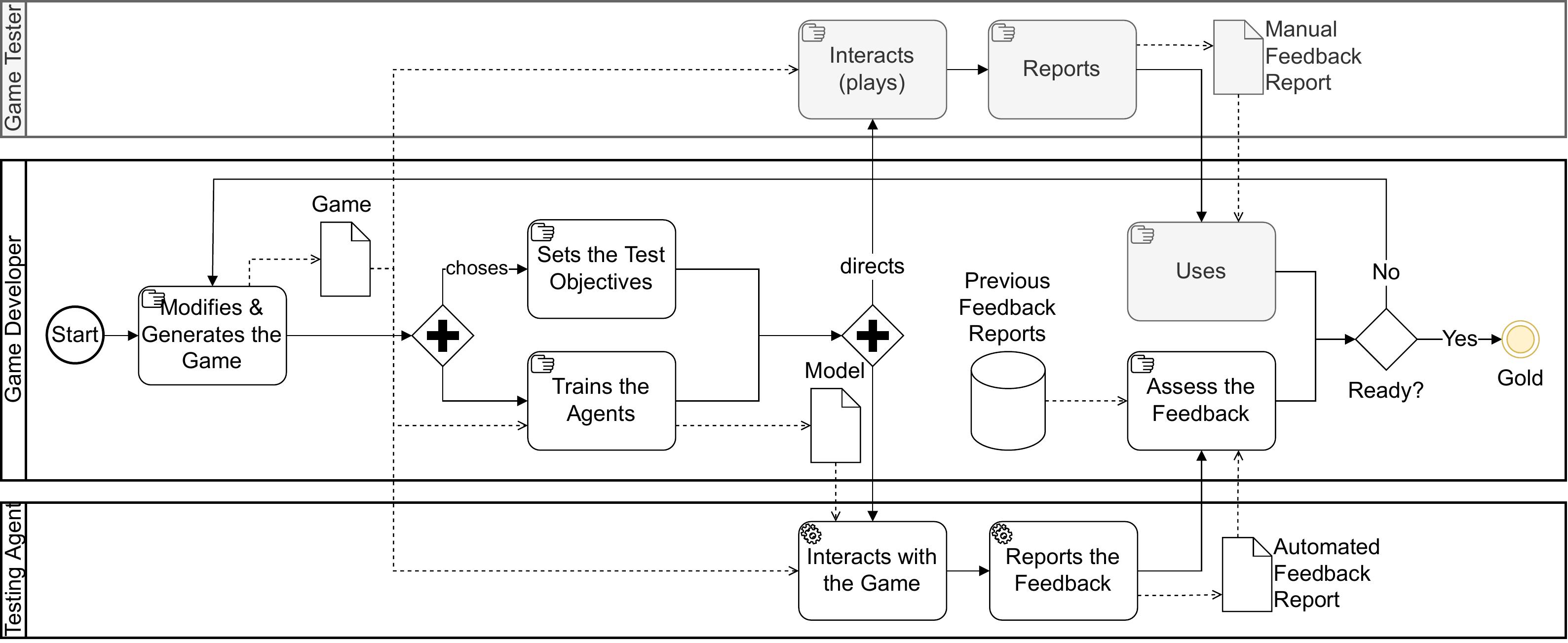}
	\caption{The testing process (in BPMN notation). The activities and artifacts show a typical manual process of game testing and how our approach complements it.}
	\label{fig:ai-test-method}
\end{figure*}

We describe, now, our approach. \autoref{fig:ai-test-method} shows the process of the feedback loop of manual game testing and how our approach automates part of it. %The \textit{Game Developer (GD)} \textit{modifies and generates} new \textit{game versions} so the \textit{Game Tester (GT)} can assess it according to the \textit{test objectives}. The game developer can then use the \textit{feedback report} to further modify the game. This is an iterative process that continues until the game is ready to be released. We complement the manual process by training the agents to autonomously play the game while measuring the metrics related to the test objective. 
There are three roles involved in this process:
The \textit{Game Developer (GD)}, which refers to Game Designers and Programmers. It is the game developer who describes and implements how things should behave in the game.
The \textit{Game Tester (GT)} (Gameplay Tester, Playtesters or QA) find bugs and any other abnormality in the game. Game testers should test game quality by verifying gameplay, logical consistency, observability, progressive thinking, reasoning ability, and exhaustively testing features, game strategy, and functionality \cite{Aleem2016a}. Therefore, game testers should understand the principles and the characteristics behind games and especially understand the game development context \cite{Santos2018}.
The \textit{Testing Agent (TA)} is the autonomous agent that interacts with the game and reports the findings according to the test objectives. 

% the video game
% Video games are composed of different elements that are linked together  \cite{Zubek2020}. A \textit{video game} is composed of \textit{game systems} which have multiple \textit{game mechanics}. Game mechanics are composed of \textit{game actions}: which are the player's input. They describe what happens on the game pieces. These actions cause transitions between \textit{game states} (the total of game pieces at a point in time); \textit{game pieces} are the game entities describing things in the game; finally, \textit{game rules} determine the outcome of player actions on game pieces. It is the game logic. For example, the scripts that add behaviour to the game pieces.

% \begin{figure}[!ht]
% 	\centering
% 	\includegraphics[width=1\linewidth]{game-class-diagram.pdf}
% 	\caption{UML class diagram abstraction of a video game and its components. Adapted from \cite{Zubek2020}.} 
% 	\label{fig:game-class-diagram}
% \end{figure}

\subsection{[GD] Modifies and Generates the Game}

The modifications in the game vary according to the game tester's feedback. For example, the game developer can simply tweak a parameter, like character speed, or introduce new gameplay mechanics, like the ability to jump. The bigger the change, the bigger the impact on the player's experience.

\subsection{[GD] Sets the Test Objectives}

The test objective depends on the modification made to the game. It is defined by the game developer and varies in scope. For example: \textit{find bugs}, \textit{explore the levels}, \textit{check the character collision}, \textit{verify if the level can be completed}, among many others.

% isolating the testing parts
As video games have a large scope
%\footnote{For example, the game \textit{Elden Ring} has hundreds of monsters and dozens of main ``bosses'' that are enemies of the player, all spread out in an open world (\url{https://en.bandainamcoent.eu/elden-ring/elden-ring}). All of them must be assessed to provide a proper experience to the player.}
, developers isolate parts of the game to test. This strategy is similar to what Rare does with the game Sea of Thieves \cite{politowskiSurveyVideoGame2021a} where they use single scripted actions to verify the object's behaviour, like opening a door, for example.

\subsection{[GD] Trains the Agents}

Playing the game is a sequential decision-making process, where the players continuously make decisions and take actions based on received observations. Therefore, this problem can be modelled as a Markov Decision Process (MDP) \cite{zhengWujiAutomaticOnline2019}.  The \textit{Agent (Player)} interacts with the \textit{Game System}. The \textit{Game State} is the observation (representation) of the Game System. What the Agent can ``see'' of the game. The Game Action is a set of possible decisions (move, attack, jump, etc). Finally, \textit{Agent's Reward} is the feedback used to measure the success  or failure of the agent's actions in achieving some goal (winning, surviving, etc). The autonomous agent (or model) is the output of the training process. It is the autonomous agent who interacts with the game.

% \begin{figure}[!ht]
% 	\centering
% 	\includegraphics[width=1\linewidth]{mdp.pdf}
% 	\caption{Markov Decision Process (MDP) and its relation with the video game. At each step (point in time), the agent (1) executes an action, (2) observes the new state, and (3) receives the reward. It then selects the best action according to the algorithm being used, like PPO or A2C for example.} 
% 	\label{fig:markov}
% \end{figure}

\subsection{[GT \& TA] Interacts with the Game}

Testing a video game means playing it \cite{politowskiSurveyVideoGame2021a}. 
%Therefore, as the complexity and scope of the game increase, more time is required to properly assess the game. Big studios have teams of testers, commonly named QA teams, to get good testing coverage of the game mechanics. Yet, small studios and indie developers do not have this luxury. Instead of the human interacting with the game, the autonomous agent will do it. 
% Typically, the game development loop is an iterative process where the developer modifies the game until it provides the experience she means to obtain. 
For every new change in the game, game testers interact with the game and provide feedback to the game developers who then change the game. This trial-and-error process relies on the empirical knowledge of the team and could be faster, more scalable, and more efficient. Indeed, as developers perform multiple changes or permutations thereof, keeping track of what works best for the game quickly becomes overwhelming. 

\subsection{[GT \& TA] Reports the Feedback}

%Game testers' reports resulting from the game testing are written in plain text with natural language. The tester specifies the issue and the steps to achieve such a state. Recreating these cases is not trivial, especially in games with random mechanics.

While the autonomous agent interacts with the game, we collect the metrics related to the testing objective. Game testers and testing agents have different testing objectives. For example, humans can assess subjective details of the game, like engagement heuristics, while autonomous agents can handle trivial details that are a burden for the human tester, like repetitive checks in the game versions.

\subsection{[GD] Assess the Feedback}

The game developer uses the feedback report to decide if the game is good enough to be released. Otherwise, they restart the process by making new modifications to the game.
After receiving the feedback containing the data about the agent's performance, they can compare it with the previous versions of the game.

%With this data, the game developer can (1) check the balance of the game between all the builds, (2) check the difficulty among the player skill level, and (3) check game design inconsistencies using random agents. 

\section{Implementation} \label{sec:implementation}

\begin{figure}[!htb]
	\centering
	\includegraphics[width=.9\linewidth]{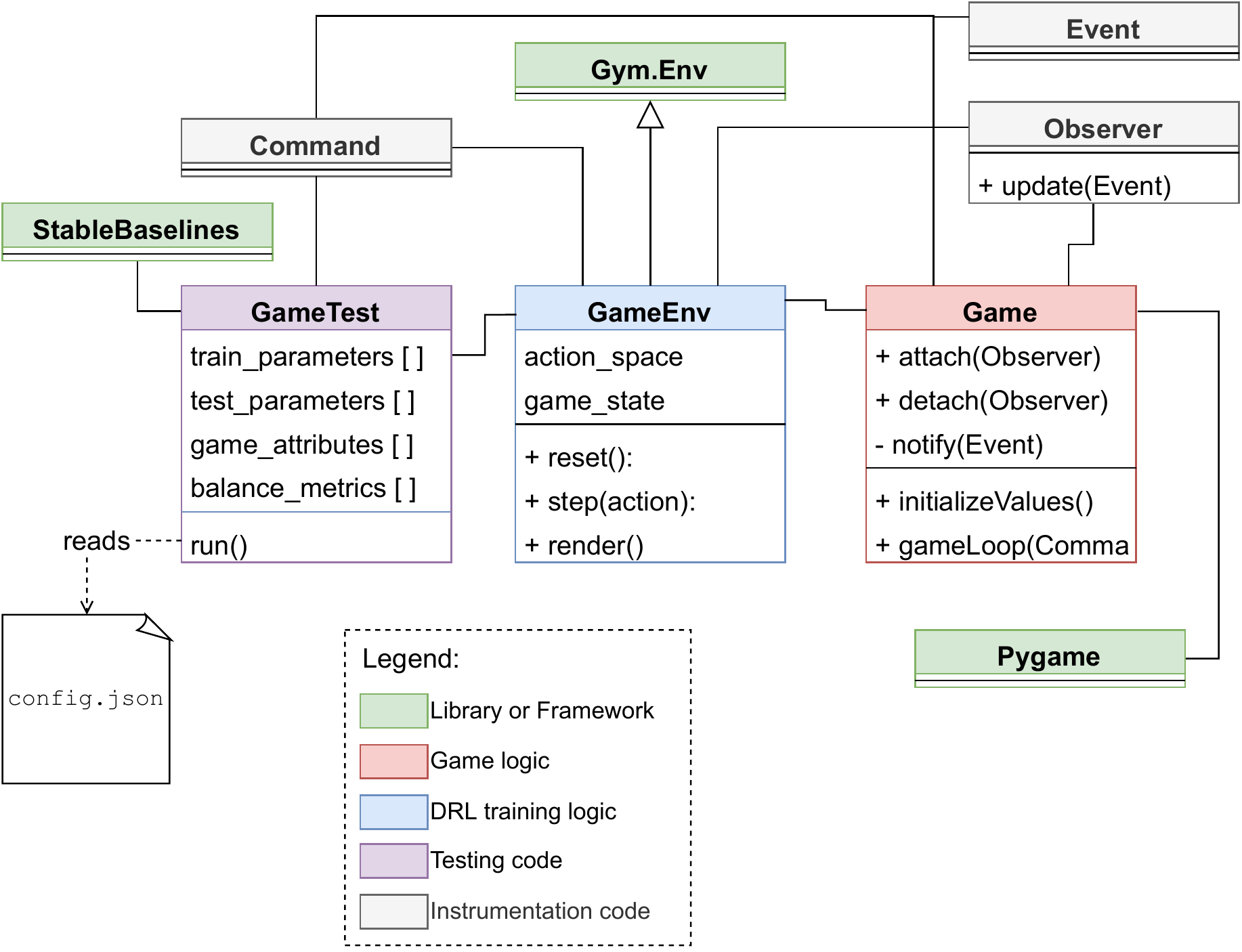}
	\caption{Architecture of our approach.}
	\label{fig:game-architecture}
\end{figure}

We now describe how we implemented our approach.
One of the issues was how to separate the concerns between the game code, the libraries and frameworks (game engine), the DRL training code, and the testing detail (number of runs, logs, etc). 
\autoref{fig:game-architecture} shows the UML-like diagram of our architecture. We separated each class and made them responsible for only one job. We started with the first element, the \textit{game logic} which is the game source code and its related assets. The game uses ``Pygame'' as \textit{framework} (game engine). Then, we used two other \textit{libraries} so we can use DLR methods to create the autonomous agents: Gym and StableBaselines. The \textit{training logic} is the element that handles the training parameters like the reward function. To link the game and training logic, we used some \textit{instrumentation} code based on the design patterns Observer and Command. Finally, we use another element to manage the \textit{testing code}.

% \subsubsection{The Games}

% For this study, we decided to focus on one type of game: \textit{Platformers}. Platformers are a video game genre where players control a game character to jump between platforms while avoiding obstacles and enemies that can kill them. They are notorious for being difficult and their designs can sometimes infuriate players rather than provide fun \cite{bostan_game_2020}. %We believe this type of game fits with our purpose of assessing the game balance. Therefore, we chose two games of the platformer genre to perform our case study on Sections \ref{sec:case-batkill} and \ref{sec:case-jungle}.

% \subsubsection{The Game Engine}

%To implement the games we use the game engine \textit{Pygame}, which is a Python framework that provides basic functions to help game development. Pygame uses the game loop structure: handling starting dealing with the \textit{inputs} from the player; \textit{updating} the game entities, which changes the game state; and \textit{rendering} the new frame of the game.

\subsection{[GD] Modifies and Generates the Game}
% \paragraph{Game Versions}
We simulate the development process by considering different versions of a game. The changes in version\#1 are carried to the further versions. 
%For some of them, we re-trained the agents because the gameplay changes were significant.
% \paragraph{The Game Difficulty} 
To have different versions of a game, we modify the game according to its game mechanics. We modify the game difficulty using the \textit{Doubling and Halving} balancing methodology \cite{schell_art_2008}. It says the developer should change the game attributes by high amounts: doubling or halving them. The objective of this method is to change something so that you can actually feel the difference right away. 

\subsection{[GD] Sets the Test Objectives}

% \subsection{Game Balance Stuff}
We focused our testing feedback on two game balance types: \textit{Challenge vs.\ Success} is about keeping the player engaged considering the game difficulty and the player's skills; \textit{Skill vs.\ Chance} is about games that depend, more or less, on luck instead of the players' skills to succeed. 
% \paragraph{Player's Skill Levels} 
To better assess the game we use game testers (humans and agents) with different skill levels: \textit{novice} and \textit{professional}. Also, we added one random agent that performs actions without any reasoning.

% \begin{figure}[!htb]
% 	\centering
% 	\includegraphics[width=1\linewidth]{difficulty-skills.pdf}
% 	\caption{The relation between game difficulty and player's skill. Adapted from \citet{schell_art_2008}. If play is too challenging, the player becomes frustrated. But if the player succeeds too easily, they can become bored. Keeping the player on the middle path means keeping the experiences of challenge and success in proper balance.}
% 	\label{fig:difficulty-skills}
% \end{figure}

\subsection{[GD] Trains the Agents}

Our approach uses DRL to train the models and create the Autonomous Agents. We use the game system to define (a) how to represent the game state and (b) what will be the Agent's Reward. These two details are specific to the game being tested and can be done in many ways. %For example, for a 2D game like Candy Crush, we could represent the game state with a matrix showing the board and reinforce the agent's behaviour by giving \texttt{+1} as a reward when the result of the action scores a point.

% \paragraph{The DRL Library}
To train the agents, we use a Python library called \textit{Gym}. It provides a number of already defined environments, which are games with defined action space and reward functions. This library is usually used to assess new machine-learning models. 

Using Gym, we defined a custom environment that includes an \textit{action space} with the commands that are possible within the game, and a \textit{reward} function. We also used \textit{Stable Baselines} library, providing a selection of machine learning models, like \textit{PPO} and \textit{A2C}, to train the agents to play the game.

% \subsubsection{Agent's Skill Level}
We divided the skill level of the agents by training time, that is, the \textit{novice} is trained with \textit{100K steps} while the \textit{professional} with \textit{1M steps}. The machine used to run the training has a CPU Core i7 2.6 GHz, A GPU NVIDIA GeForce RTX3070, 32 GB DDR4 or RAM, and an SSD hard drive. As for the software, we use the Python language with the Stable Baseline library running on Windows with NVIDIA CUDA.

% training parameters
To train the autonomous agents we used \textit{Training Parameters} (\autoref{tab:training-params}). For our experiments we chose to use the model-free DRL \texttt{models} because (1) in our game environment, we cannot predict state transitions and rewards, and (2) the training cost (computation time) is lower. Thus, we chose two different models to train the agents: \textit{PPO} and \textit{A2C}. The \texttt{action space}, \texttt{reward function}, and \texttt{observation space} vary from game to game.

\begin{table}[!htb]
	\footnotesize
	\centering
	\caption{The training parameters used to train the autonomous agents.}
	\label{tab:training-params}
	\begin{tabularx}{\linewidth}{@{}llX@{}}
		\toprule
		Params & Type/Value & Description \\ \midrule
		\texttt{train} & Boolean & Re-train or not the agents \\
		\texttt{model} & String & The machine learning model used to train the agents \\
		\texttt{action space} & Array & String indicating the action of the game (LEFT, ATTACK, etc) \\
		\texttt{reward function} & Float & Values, positives and negatives, defined by an heuristic \\
		\texttt{observation space} & Array & The state of the game, what the agent knows and ``see'' \\ \bottomrule
	\end{tabularx}
\end{table}

% Retraining the Agents
Depending on the modification of the game, it is necessary to re-train the models (agents) for each new version. For example, when a new game mechanics heavily modifies the gameplay. 

% \paragraph{Validating the Agents with Human Testers}
We used human players to validate the performance of the agents. The agent must have a similar performance compared to the human players. Also, the performance across different versions of the game should follow the same pattern/trend.

\subsection{[TA] Interacts with the Game}

% \paragraph{Play time}
We defined a play session of 180 seconds where the autonomous agent plays the game two times. We use the median of the two runs to calculate each performance.

% testing parameters
\autoref{tab:test-params} shows the \textit{Testing Parameters} used in the game test scenarios. The player plays the game for \textit{N} seconds, \textit{M} times. Either a human or an agent plays the game. Each one is assigned a skill level, with the exception of the random agent. The human professional is someone with gaming experience while the novice is not a regular game player. 

\begin{table}[!htb]
\centering
\footnotesize
\caption{Testing parameters.}
\label{tab:test-params}
\begin{tabular}{@{}lll@{}}
    \toprule
    Parameter & Type/Value & Description \\ \midrule
    \texttt{time} & Integer & The time to be played in seconds \\
    \texttt{run} & Integer & Number of runs \\
    \texttt{session} & \{human, ai-play, random\} & The player of the session \\
    \texttt{skill} & \{novice, professional\} & The skill level of the player \\
    \texttt{version} & String & The game versions \\
    % \texttt{train} & Boolean & If the AI will re-train the agents \\
    % \texttt{model} & \{ppo, a2c\} & The machine learning model used to train the agents\\
    \bottomrule
\end{tabular}
\end{table}

\subsection{[TA] Reports the Feedback}

As we are tackling the issue of balancing the game, we created metrics related to the score of the agent playing the game. These metrics are variables that work as a proxy for the agent's performance and, therefore, the game balance. They vary for each game but, in general, are related to the score of the game. 

\subsection{[GD] Assess the Feedback}

We compare the balance metrics across the different versions of the game:

\begin{itemize}
    \item To balance the \textit{Challenge vs.\ Success} we check spikes on the balance metrics when the agent plays in each game version. Also, we compare the performance of novice and professional skill levels.
    \item To balance the \textit{Skill~vs.~Chance} we compare the performance of the random agent with the other trained agents. If the random agent performs better, the game has a balance problem.
\end{itemize}

%%%
% CASE A BATKILL
%%%

\section{Case Study A. Batkill} \label{sec:case-batkill}

% \subsection{About the Game}

For Case Study A, we modified a 2D action platformer open-source game called Batkill\footnote{\url{https://github.com/python-aficionado/batkill}}. We did not touch the rules of the game. 
It consists of a single screen, where the character tries to stay alive while bats fly toward him. The goal is to kill as many bats as possible without being hit. For each bat killed, the player gets one point (\texttt{+1 score}). The player loses one life for each hit (\texttt{-1 life}). The bats spawn faster as the players kill them. The bats spawn in random locations. The actions are \texttt{LEFT, RIGHT, ATTACK, JUMP}. 
% \autoref{fig:batkill-states} shows the game and its actions.

% \begin{figure}[!htb]
% 	\centering
% 	\includegraphics[width=1\linewidth]{actions.png}
% 	\caption{The character actions of the game Batkill. From bottom to top: \textit{standing}, \textit{attacking}, \textit{running}, and \textit{jumping}.}
% 	\label{fig:batkill-states}
% \end{figure}

% \subsection{Manual Game Testing}

\subsection{[GD] Modifies and Generates the Game} \label{sec:case-a-step1}

We created five different versions of the game Batkill (\autoref{tab:game-versions-batkill}) by changing a set of variables in the game:
\texttt{bats} is the number of enemies on screen;
\texttt{bat\_speed} is the enemies' movement speed;
\texttt{attack\_cooldown} is the time between the character's attacks;
\texttt{jump} if the character can jump or not.
We expect an increase in the difficulty on versions \#2, \#3, and \#4. On version \#5, with the addition of the jump mechanic, the difficulty should decrease as the player has one more resource to avoid enemies.

\begin{table}[!htb]
	\centering
    \footnotesize
	\caption{The game versions for the game Batkill.}
	\label{tab:game-versions-batkill}
	\begin{tabular}{@{}lrrrl@{}}
		\toprule
		Version & \texttt{bats} & \texttt{bat\_speed} & \texttt{attack\_cooldown} & \texttt{jump} \\ \midrule
		\#1 & 2 & 3 & 10 & FALSE  \\
		\#2 & 3 & 6 & 10 & FALSE  \\
		\#3 & 3 & 6 & 10 & FALSE  \\
		\#4 & 3 & 6 & 15 & FALSE  \\
		\#5 & 3 & 6 & 15 & TRUE  \\ \bottomrule
	\end{tabular}
\end{table}

\subsection{[GD] Sets the Test Objectives}

To assess the balance of the game Batkill, we measure two variables. \texttt{bats\_killed}, which is the number of enemies killed by the character; and \texttt{hits\_taken}, which is the number of times an enemy hits the character. To have a balance between kills and hits, the balance metric (\texttt{score}) for the game Batkill is given by: $score = bats\_killed - hits\_taken$.

% \begin{equation}
%     score = bats\_killed - hits\_taken
% \end{equation}

% \subsection{Automated Game Testing}

\subsection{[GD] Trains the Agents}

% reward
To train the agents to play the game Batkill, we started with giving rewards when an enemy is killed (\texttt{BAT\_KILLED +5}) or a hit was taken (\texttt{HIT\_TAKEN -5}). After that, we used a small negative reward (\texttt{ATTACK -0.1}) to avoid making the agent use the attack for no reason. After seeing humans playing the game, we verified that they did not jump often. That is why we penalized the agent if he jumped (\texttt{JUMP -0.2}). The idea is to keep the character on the ground for more time. We also gave a small reward when the player moved toward the enemy (\texttt{MOVING\_TOWARDS +0.1}). Finally, we gave a small reward if the agent faced the nearest enemy (\texttt{FACING\_NEAREST\_BAT +0.2}). That is a movement humans do naturally, and we try to hint here.

% For the observation space, the state of the game, that is, what the agent knows and ``see'', we defined this set of variables:

% \begin{itemize}
% 	\item \texttt{player\_x}: the character horizontal position.
% 	\item \texttt{player\_y}: the character vertical position.
% 	\item \texttt{player\_direction}: the direction the character is facing (left or right).
% 	\item \texttt{player\_facing\_bat}: if the player is facing the bat.
% 	\item \texttt{player\_attack}: if the player attack (True or False).
% 	\item \texttt{player\_cooldown}: the time between the attacks.
% 	\item \texttt{bat\_alive}: if the bat is alive.
% 	\item \texttt{bat\_direction}: the direction the bat is facing (left or right)
% 	\item \texttt{bat\_x}: the bat horizontal position.
% 	\item \texttt{bat\_speed}: the bat speed.
% 	\item \texttt{bat\_distance\_to\_player}: the bat distance to the player.
% 	\item \texttt{bat\_in\_attack\_range}: if the bat is in attack range.
% \end{itemize}

The PPO model had better performance (bigger reward) in all versions when the steps were more than 100K. For the PPO model, training a ``novice'' agent took around six minutes, while the ``professional'' took around one hour. The A2C model took about 10\% more time to do the same training process. 

% \begin{figure}[!t]
% 	\centering	
% 	\subfloat[Skill level: Novice (100K steps of training).]{
% 		\includegraphics[width=1\linewidth]{rew_mean_novice_builds.pdf}
% 		\label{fig:train-results-novice}
% 	}		
% 	\subfloat[Skill level: Professional (1M steps of training).]{
% 		\includegraphics[width=1\linewidth]{rew_mean_pro_builds.pdf}
% 		\label{fig:train-results-pro}
% 	}	
% 	\caption{Training results for the game Batkill.}
% 	\label{fig:train-graphs}
% \end{figure}

\subsection{[TA] Interacts with the Game}

The trained agent interacts (plays) with the game autonomously. The behaviour of the agent resembles a human playing the game. However, the agent performs movements that are uncanny for humans. That is, a human player would never play like that. 

For example, even after putting a penalty for it in the reward functions, the agent used \textit{jump} and \textit{attack} constantly whether the enemy was near or not. Nevertheless, there were moments the agent stayed still, like a human player.

\subsection{[TA] Reports the Feedback}

The \autoref{tab:score-batkill} shows the results of the feedback report for the PPO and A2C agents, as well as the Random agent and the Human player. The score of each agent is also separated by skill level. The negative values mean that the agent/player got hit more times than kills.

\begin{table}[!ht]
\footnotesize
\centering
\caption{Median of \texttt{SCORE (bats\_killed - hits\_taken)} for the game Batkill.}
\label{tab:score-batkill}
\begin{tabular}{@{}lrr|rr|rr|r@{}}
\toprule
Version &
  \multicolumn{2}{l|}{Human} &
  \multicolumn{2}{l|}{Agent PPO} &
  \multicolumn{2}{l|}{Agent A2C} &
  \multicolumn{1}{l}{Random} \\ \cmidrule(l){2-8} 
 &
  \multicolumn{1}{l}{Pro} &
  \multicolumn{1}{l|}{Novice} &
  \multicolumn{1}{l}{Pro} &
  \multicolumn{1}{l|}{Novice} &
  \multicolumn{1}{l}{Pro} &
  \multicolumn{1}{l|}{Novice} &
  \multicolumn{1}{l}{} \\ \midrule
\#1 & 78  & 59  & 18  & 23  & 29   & 13   & -13 \\
\#2 & 21  & 6   & -7  & 7   & -44  & -47  & -27 \\
\#3 & -67 & -86 & -53 & -63 & -112 & -122 & -73 \\
\#4 & -74 & -92 & -96 & -86 & -121 & -123 & -98 \\
\#5 & -36 & -1  & -40 & -47 & -56  & -51  & -56 \\ \bottomrule
\end{tabular}
\end{table}

% On version\#2, the human player killed more enemies, because of the one extra bat added. However, there is a big spike in hits taken. When we increase the bat speed on version\#3, the player had another spike in hits taken and fewer bats killed. The worst score was in version\#4 when we increase the time between the character's attacks. Finally, when we add jump on version\#5 the player could avoid damage, similar to what he had on version\#2, but killed fewer bats. This is because of the play style, either you play aggressively attacking more than avoiding damage, or you chose to jump and attack when it is safe.

% The human novice got better results on version\#5 when we compare it with the human professional. However, as \autoref{fig:res-batkill-novice-vs-pro} shows, the novice had fewer kills and fewer hits on all builds. On build\#5, the novice human killed less but took much fewer hits too. This is because the jump allows the player to choose between playing carefully and more aggressively.

% \begin{figure}[!htb]
% 	\centering
% 	\includegraphics[width=1\linewidth]{res-batkill-human-novice-vs-pro.pdf}
% 	\caption{Bats killed and Hits taken for Human novice and professional on Case Study A. Batkill.}
% 	\label{fig:res-batkill-novice-vs-pro}
% \end{figure}

\subsection{[GD] Assess the Feedback}

\begin{figure}[!htb]
	\centering
	\includegraphics[width=.9\linewidth]{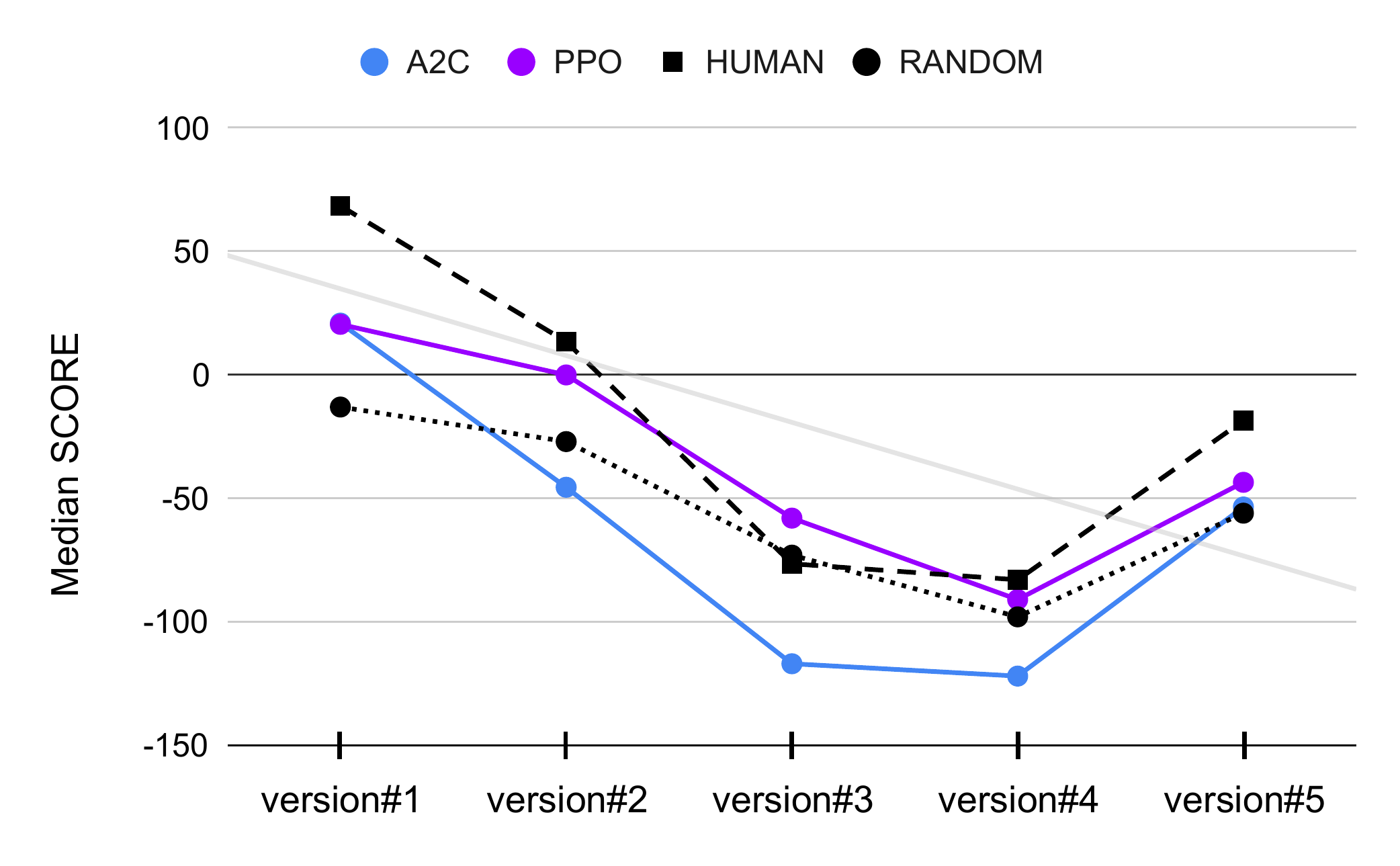}
	\caption{Results of all agents on Case Study A. Batkill.}
	\label{fig:results-batkill}
\end{figure}

% overall of the game difficulty
\autoref{fig:results-batkill} shows the results of the humans and agents playing the game. The \texttt{score} is the average between novice and professional players/agents. Version\#1 is the easiest because the agents and human players had the best score among all other versions. Version\#4 is the hardest, followed by versions \#3, \#5, and \#2.

% show that human and ppo are similar
In all game versions, the human outperforms any other autonomous agent, except on version\#3. The PPO agent is the one that plays similarly to the human player, as the score across the game versions follows a similar pattern. In general, the A2C agent results are the lowest across all versions, except in the first version.

% PPO is the agent that plays more aggressively, taking more hits and killing more bats. On the other hand, A2C took more cautious actions, especially on build 5.

% Challenge vs.\ Success: spikes in difficulty
\subsubsection{Challenge vs.\ Success}
We can identify three big spikes in difficulty. The first two spikes made the game harder while the last one made the game easier:

\begin{itemize}
    \item \textbf{Difficulty Spike}: on game version\#2, when we increase the number of enemies, the game got \textit{harder}.
    \item \textbf{Difficulty Spike}: on version\#3, when we increase the speed of the enemies, the game got \textit{harder}.
    \item \textbf{Difficulty Spike}: on version\#5, when we added the ``jump'' mechanic, the game got \textit{easier}.
\end{itemize}

% Skill~vs.~Chance: compare PPO with random
\subsubsection{Skill vs.\ Chance}
We identified two game versions where luck was more important than the player's skill. In those cases the random agent had similar (or better) performance to the other agents and human players:

\begin{itemize}
    \item \textbf{Chance}: on version\#3, when we increase the number of enemies, the game depends more on \textit{luck} than skill.
    \item \textbf{Chance}: on version\#4, when we decreased the time between the character attacks, the game depends more on \textit{luck} than skill.
\end{itemize}

% This means that the game, in that state (3 bats with more speed), does not rely on the player's skill but on ``button-mashing'' techniques. Therefore, builds\#3 seems to be the worst game state in terms of balancing. 

\begin{myboxii}[Summary - Case Study A. Batkill]	
\begin{itemize}
    \item The PPO agent is the one that follows a similar pattern as the human players.
    \item Across the game versions, we found two spikes of difficulty that made the game harder and one spike that made it easier.
    \item The performance of the random agent on versions \#3 and \#4, compared to the other agents, shows that these two versions do not favour the player's skill. However, the introduction of jumping on version\#5 helped making the game more skill-based.
    % \item The addition of a new gameplay feature (the ability to jump), shows that humans can take different strategies when playing (aggressive or caution in our case). This hints to the importance of using multiple balance metrics to measure the game.
    % \item Compared to PPO, A2C agents got better rewards during the novice training (100K steps) but worst results when the agent plays the game.
\end{itemize}	
\end{myboxii}

%%%
% CASE B JUNGLE
%%%

\section{Case Study B. Jungle Climb} \label{sec:case-jungle}

We modified an open-source, 2D ``infinite-runner'' platform game\footnote{\pageClimb} called Jungle Climb\footnote{\url{https://github.com/elibroftw/jungle-climb}}. It consists of a single screen, where several platforms are drawn. Each platform has gaps randomly generated. The player character is initially placed below the platforms and has to climb them by jumping between gaps. After passing through the second platform, the screen will start scrolling upward. The objective of the player is to keep the character below the bottom line of the screen as long as possible, not letting the player be ``scrolled down'' for too long with the screen.

For each step, the player is able to survive while above the second platform, it gets one point (\texttt{+1 point}). The scrolling speed of the platform increases as time passes. The actions are \texttt{LEFT}, \texttt{RIGHT} and \texttt{JUMP}. 
% \autoref{fig:jungle-states} shows the game and its actions. 

% \begin{figure}[!htb]
% 	\centering
% 	\includegraphics[width=1\linewidth]{actions-jungle.png}
% 	\caption{The character actions of the game Jungle Climb. From the bottom to up: \textit{standing}, \textit{jumping}, and \textit{running}.}
% 	\label{fig:jungle-states}
% \end{figure}

\subsection{[GD] Modifies and Generates the Game}

We created three different versions of the game Jungle Climb (\autoref{tab:game-versions-jungle}) by changing two variables in the game:
\texttt{shift\_speed} is the rate the screen scrolls up; and \texttt{max\_gaps}, which is the maximum number of gaps in each platform. We expect version\#2 to increase the difficulty of the game and version\#3 to make the game easier.

\begin{table}[!htb]
	\centering
    \footnotesize
	\caption{The game versions - Jungle Climb}
	\label{tab:game-versions-jungle}
	\begin{tabular}{@{}lrr@{}}
		\toprule
		Version & \texttt{shift\_speed} & \texttt{max\_gaps} \\ \midrule
		\#1   & 1           & 1           \\
		\#2   & 2           & 1          \\
		\#3   & 2           & 2           \\ \bottomrule
	\end{tabular}
\end{table}

\subsection{[GD] Sets the Test Objectives}

To assess the balance of the game Jungle Climb, we measure two variables: \texttt{max\_points}, which is the maximum number of points that shows how long the character was alive; and \texttt{max\_correct\_jumps}, which is the maximum of correct jumps, that is, when the jump connects to the next platform. The \texttt{score} is given by the rate between points and correct jumps: $score =  max\_points + (max\_correct\_jumps * 100)$.

% \begin{equation}
% score =  max\_points + (max\_correct\_jumps * 100)
% \end{equation}

\subsection{[GD] Trains the Agents}

To train the game Jungle Climb we use the following set of rewards. At the beginning of every step, we compute the \texttt{Below Threshold Reward (BTR)}. If the character has not passed the threshold of the screen where it starts to shift, the score will always be zero and the BTR will be greater than zero. We use this value so we can give the agent different motivations so it can exit this initial state more quickly: \texttt{BTR = time\_elapsed * 5 if score == 0 else 0}.

% We check if the character is not under the gap and, if the distance between the character and the gap is decreasing, we give \texttt{+100} and  \texttt{-100} otherwise.

% If the character is under the first gap, we check first if he is facing the second gap (in the upper row), and give \texttt{+100}. If the character's position on Y-axis is decreasing, the reward is \texttt{+100 + BTR}, otherwise, \texttt{-100 - BTR}. Finally, we give a penalty to the agent so he can get out of the first platform. Thus, if the \texttt{BTR} is bigger than zero and the character repeats the previous step, we give \texttt{-100}.

\subsection{[TA] Interacts with the Game}

The trained model (agent) interacts (plays) with the game autonomously. However, aside from our efforts to train the model, the behaviour of the agent does not resemble a human player. The agent jumps continuously and, most of the time, without purpose. 

\subsection{[TA] Reports the Feedback}

The \autoref{tab:score-jungle} shows the results of the feedback report for the PPO and A2C agents, as well as the Random agent and the Human player. The score is separated by skill level.

\begin{table}[!htb]
\footnotesize
\caption{Median of \texttt{SCORE} for the game Jungle Climb.}
\label{tab:score-jungle}
\begin{tabular}{@{}lrrrrrrr@{}}
\toprule
Version &
  \multicolumn{2}{l|}{Human} &
  \multicolumn{2}{l|}{Agent PPO} &
  \multicolumn{2}{l|}{Agent A2C} &
  \multicolumn{1}{l}{Random} \\ \cmidrule(lr){2-7}
 &
  \multicolumn{1}{l}{Pro} &
  \multicolumn{1}{l|}{Novice} &
  \multicolumn{1}{l}{Pro} &
  \multicolumn{1}{l|}{Novice} &
  \multicolumn{1}{l}{Pro} &
  \multicolumn{1}{l|}{Novice} &
  \multicolumn{1}{l}{} \\ \midrule
\#1 & 3262 & 2890 & 3597 & 1576 & 788 & 407 & 1371 \\
\#2 & 1908 & 1712 & 2251 & 850  & 519 & 305 & 683  \\
\#3 & 1885 & 1591 & 2154 & 1102 & 100 & 338 & 615  \\ \bottomrule
\end{tabular}
\end{table}

\subsection{[GD] Assess the Feedback}

\begin{figure}[!htb]
	\centering
	\includegraphics[width=.9\linewidth]{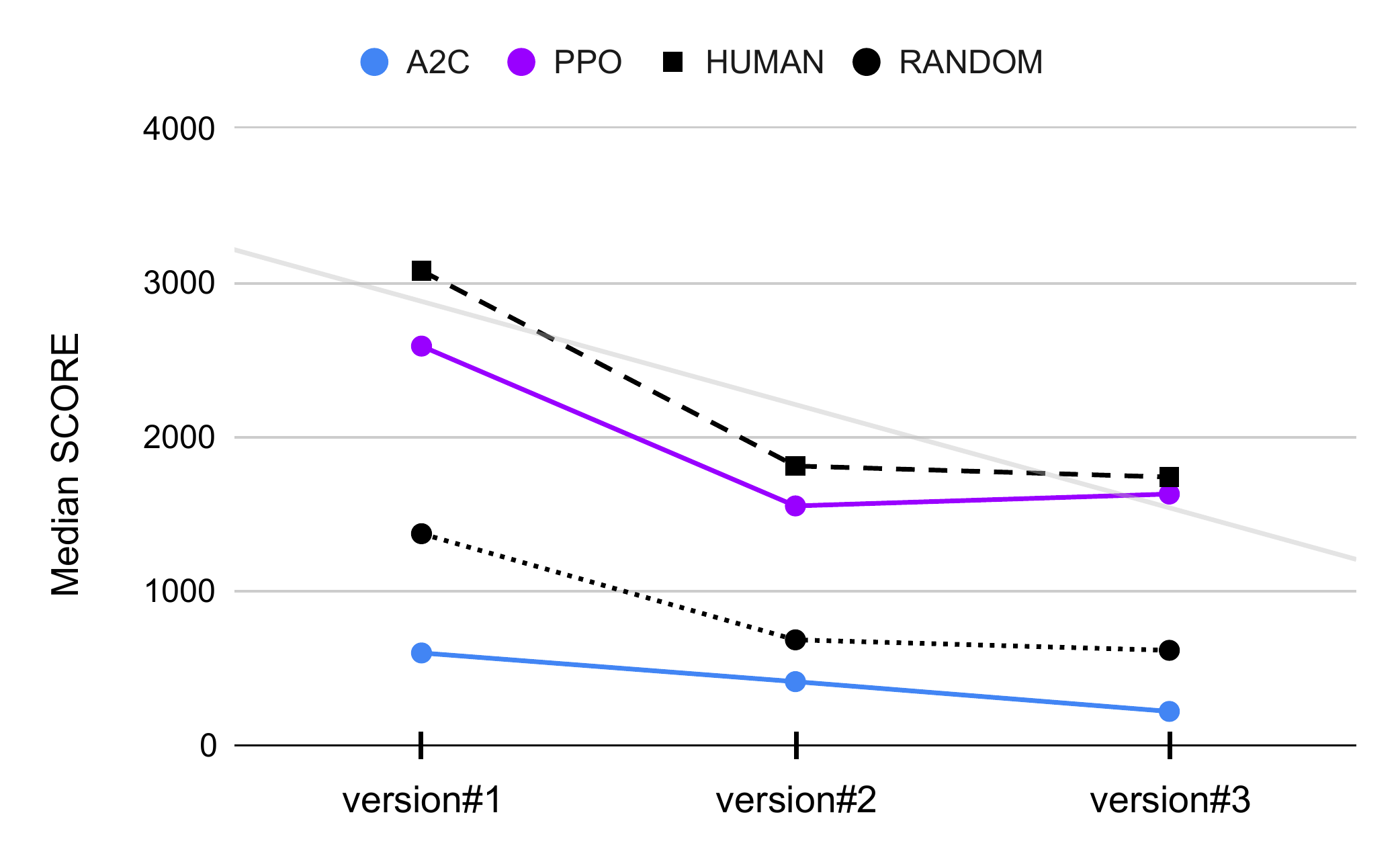}
	\caption{Results of all agents on Case Study B. Jungle Climb.}
	\label{fig:results-jungle}
\end{figure}

\autoref{fig:results-jungle} shows the results (score) of all the agents playing the game versions. Version \#1 is the easiest to play. The difficulty across the versions increases on version\#2 and maintains on version\#3. Even by adding one more gap to jump into each platform.

The curve for the PPO agent and Human players are very similar, which shows that this agent is a good proxy to mimic human players.
The PPO agent outperforms, in all three versions, all the other agents. On the other hand, the A2C agent performs poorly, even when compared to the random agent. 

\subsubsection{Challenge vs.\ Success}
We could identify one big spike in difficulty:

\begin{itemize}
    \item \textbf{Difficulty Spike}: on game version\#2, when we increase the speed the screen shifts, the game got \textit{harder}.
\end{itemize}

% Skill~vs.~Chance: compare PPO with random
\subsubsection{Skill vs.\ Chance}
The random agent has a similar pattern to the human player but with a very low score across all game versions. This indicates that the game does reward the player's skill in all versions. Therefore:

\begin{itemize}
    \item \textbf{Skill}: on versions \#1, \#2, and \#3,  the game depends more on \textit{skill} than luck.
\end{itemize}

\begin{myboxii}[Summary - Case Study B. Jungle Climb]
	
\begin{itemize}
    \item The performance pattern of the PPO agent is similar to the human players.
    \item There is a spike in difficulty on version\#2 that is not balanced on version\#3 by adding more gaps to jump. 
    \item The random agent does not outperform the human or the PPO, showing that the game is skill-based.
\end{itemize}
	
\end{myboxii}

%%%
% DISCUSSION
%%%

\section{Discussion} \label{sec:discussion}

\subsection{Balancing Challenge vs. Success}

To deliver a good experience to the players, keeping them away from boredom or anxiety, game development requires trial and error and much experimentation. Our results show that it is possible to systematically automate part of the balance testing process. In our case studies, we showed that the agents could mimic the player's achievements and struggles, even without re-training the models. Thus we showed that it is possible to rely on autonomous agents to proxy the human's skill levels. These agents identified the spikes in difficulties among the versions. This setup provided quick feedback that game developers can use to promote changes in the game design. 

\subsection{Balancing Skill vs. Chance }

% \cris{For a game that demands player skill, the random agent must be lower than the agents/human.}

Games are complex systems that mix mechanics that demand both skill and luck (chance). Finding the ``sweet spot'' requires a subjective aspect that only experienced developers have. Using the random agents proved to be useful in spotting when the game is rewarding, or not, the player's skills. When compared to the trained agents and humans, we noticed that, in certain versions, the random agents got the best results. With a closer look at the metrics, these exact same versions were the most difficult to play. The game developers can use the information to tweak the game accordingly, that is, make the game more skill-based or chance-based.

\subsection{Expectation vs. Reality} 

With Case Study A we confirmed our design choices (Section~\ref{sec:case-a-step1}) that adding the jump ability would reduce the game difficulty, as it would add a new way to avoid getting hit. However, in Case Study B, we assumed that adding a new gap in the platforms would facilitate the climbing, which was not the case. This shows that the game developer's assumptions must be properly tested in practice. Our approach helps this process.

\subsection{Training the Agents and Balance Issues}

Training the agents demands an initial effort that pays off later in development. As each game has different mechanics, it takes time to discover which rewards will result in an agent playing the game well, or close to the human performance. While trying different rewards, the game shows some of its balance issues. For example, even a simple game like in Case Study B. Jungle Climb requires an effort to make the agents play reasonably well.

\subsection{Game Testers and Effort}

In our case studies, we asked two people to play the game for three minutes for each version, twice. Even with these two simple game scenarios and a few minutes of gameplay, both of the players reported tiredness after playing each game. This adds to the fact that manual testing on video games is tiresome and demands full concentration. In the game industry, game testers work for hours every day on the same game checking the same issues multiple times. The automation in the game testing aid these testers as well, so they can focus on subjective details of the game.

\subsection{Deterministic vs. Stochastic}

The two games used in our experiments are stochastic, they contain random gameplay elements that are not possible to predict. e.g., the bats in Case Study A and the gaps in Case Study B. Games like these pose difficulties when training the agents to play the game. As randomness is a feature in the games, automating the testing is even more valuable for these cases.

\subsection{Cost of Creating the Reward Function}

To make the DRL agents play the game, we must give rewards (or penalties). However, finding a combination of rewards that makes the agents play the game properly requires trial and error. Ultimately, the cost to produce ``heuristics'' to reward agents should be lower than creating a simple script that performs actions to the game. Thus, the process of creating cost-effective reward functions is an open challenge yet.

\section{Threats to Validity} \label{sec:threats}

We faced issues during the implementation of our approach that can become obstacles to its adoption. For example, creating the testing architecture (\autoref{fig:game-architecture}) demands an initial effort. Although this investment pays in the long run, we understand that spending engineering time with tooling instead of the game itself might not be welcomed by developers. However, many libraries allow a rapid configuration of the environment. In our cases, we used Python libraries (Gym and Stable Baselines) to speed up the process of training.

Another issue is to define reward functions so that the agents can play the game properly. Games have different mechanics and even games within the same genre, e.g. platformers, do not have common rewards. Yet, scenarios within the same game can share the same rewards. For example, in larger games, we could split the chunks of the scenarios the developer wants to test and reuse the rewards with fewer modifications.

\section{Conclusion} \label{sec:conclusion}

% conclusion implementation
In this paper, we proposed and implemented our approach to automate game testing to balance video games using autonomous agents. We described the process of training the agents, playing the game, and assessing the game balance. 
We focused on two game balance types: \textit{Challenge vs.\ Success}  and \textit{Skill vs.\ Chance}. 
We validated our testing process with two platform games. We found difficulties spikes in the game versions and identified the versions which demanded more (or less) player skill.

Game development often relies more on the developer's ``feeling'' rather than on any game specification. The result is an empirical manual cycle of development and testing. Although replacing manual testing is not possible, game developers can adopt a development pipeline with automated testing that provides quick feedback about the game balance. 

%\begin{myboxii}
Our approach brings a systematic and autonomous way to identify if the game is balanced by (1) checking the difficulty spikes and (2) if the game demands more skill or not.
% We conclude that the use of autonomous to test games is faster than the manual feedback loop and provides a viable solution for game balancing, showing spikes in difficulty between game versions and issues with the game design. 
We believe our approach is a step toward providing steady game development and games with better quality.
% This automated process must help game developers by (1) providing fast feedback about the game balance and (2) testing multiple game versions at once.
%\end{myboxii}

% \subsection{Future Work}

For future work we want to expand the scope of the experiments and find easier ways to train the agents. We also want to add more skill levels for the agents aside from different ways to play the game, that is, a player profile/persona. %For example, creating an agent that only focuses on finishing the game while others explore the states of the game. Moreover, we want to add different game genres (racing, fighting, sports, etc) and game types (3D, isometric, etc).

% Second, we only explore two types of balance problems, there are others that also require the developer's attention \cite{schell_art_2008}. For example, the \textit{Fairness} in asymmetrical games or \textit{Simple vs Complex} game mechanics. To test these subjective items it might be necessary to create a more complex testing pipeline and more ``intelligent'' agents. 

% By creating an ecosystem for automated game testing, it is possible to add other testing objectives to the balancing test, like an exploration of the game and checking its performance (in terms of frames-per-second). These metrics, together with the existent balancing ones, can provide better feedback to the game developer.

% \begin{acks}
% The authors were partially supported by the NSERC Discovery Grant and Canada Research Chairs programs.
% \end{acks}

% \bibliographystyle{abbrvnat}
% \bibliography{main.bib}
\printbibliography

@inproceedings{preussIntegratedBalancingRTS2018,
  title = {Integrated {{Balancing}} of an {{RTS Game}}: {{Case Study}} and {{Toolbox Refinement}}},
  shorttitle = {Integrated {{Balancing}} of an {{RTS Game}}},
  booktitle = {2018 {{IEEE Conference}} on {{Computational Intelligence}} and {{Games}} ({{CIG}})},
  author = {Preuss, Mike and Pfeiffer, Thomas and Volz, Vanessa and Pflanzl, Nicolas},
  year = {2018},
  doi = {10.1109/CIG.2018.8490426},
}

@inproceedings{morosanLessonsTestingEvolutionary2018,
  title = {Lessons from {{Testing}} an {{Evolutionary Automated Game Balancer}} in {{Industry}}},
  booktitle = {2018 {{IEEE Games}}, {{Entertainment}}, {{Media Conference}} ({{GEM}})},
  author = {Morosan, Mihail and Poli, Riccardo},
  year = {2018},
  doi = {10.1109/GEM.2018.8516447},
}

@incollection{morosanAutomatedGameBalancing2017,
  title = {Automated {{Game Balancing}} in {{Ms PacMan}} and {{StarCraft Using Evolutionary Algorithms}}},
  booktitle = {Applications of {{Evolutionary Computation}}},
  author = {Morosan, Mihail and Poli, Riccardo},
  editor = {Squillero, Giovanni and Sim, Kevin},
  year = {2017},
  volume = {10199},
  publisher = {{Springer International Publishing}},
   doi = {10.1007/978-3-319-55849-3_25},
}

@inproceedings{liuAutomaticGenerationTower2019,
  title = {Automatic Generation of Tower Defense Levels Using {{PCG}}},
  booktitle = {Proceedings of the 14th {{International Conference}} on the {{Foundations}} of {{Digital Games}}},
  author = {Liu, Simon and Chaoran, Li and Yue, Li and Heng, Ma and Xiao, Hou and Yiming, Shen and Licong, Wang and Ze, Chen and Xianghao, Guo and Hengtong, Lu and Yu, Du and Qinting, Tang},
  year = {2019},
  publisher = {{ACM}},
  doi = {10.1145/3337722.3337723},
}

@book{schell_art_2008,
	address = {Amsterdam ; Boston},
	title = {The art of game design: a book of lenses},
	isbn = {978-0-12-369496-6},
	shorttitle = {The art of game design},
	language = {en},
	publisher = {Elsevier/Morgan Kaufmann},
	author = {Schell, Jesse},
	year = {2008},
}

@ARTICLE{Pfau2022Dungeons2,  
	author={Pfau, Johannes and Liapis, Antonios and Yannakakis, Georgios N. and Malaka, Rainer},  journal={IEEE Transactions on Games},   
	title={Dungeons  amp; Replicants II: Automated Game Balancing Across Multiple Difficulty Dimensions via Deep Player Behavior Modeling},   
	year={2022},  
	doi={10.1109/TG.2022.3167728}
}

@misc{mnih2013playing,
	title={Playing Atari with Deep Reinforcement Learning}, 
	author={Volodymyr Mnih and Koray Kavukcuoglu and David Silver and Alex Graves and Ioannis Antonoglou and Daan Wierstra and Martin Riedmiller},
	year={2013},
}

@article{Justesen2020,
	title = {Deep {Learning} for {Video} {Game} {Playing}},
	doi = {10.1109/TG.2019.2896986},
	journal = {IEEE Transactions on Games},
	author = {Justesen, N. and Bontrager, P. and Togelius, J. and Risi, S.},
	year = {2020},
}

@Article{Aleem2016a,
	author  = {Aleem, Saiqa and Capretz, Luiz Fernando and Ahmed, Faheem},
	title   = {{Critical Success Factors to Improve the Game Development Process from a Developer's Perspective}},
	journal = {Journal of Computer Science and Technology},
	year    = {2016},
	doi     = {10.1007/s11390-016-1673-z},
}

@InProceedings{politowskiSurveyVideoGame2021a,
	author    = {Politowski, Cristiano and Petrillo, Fabio and Gueheneuc, Yann-Gael},
	booktitle = {2021 {{IEEE}}/{{ACM International Conference}} on {{Automation}} of {{Software Test}} ({{AST}})},
	title     = {A {{Survey}} of {{Video Game Testing}}},
	year      = {2021},
	doi       = {10.1109/AST52587.2021.00018},

}

@InProceedings{Santos2018,
	author    = {Ronnie E. S. Santos and Cleyton V. C. Magalhaes and Luiz Fernando Capretz and Jorge S. Correia-Neto and Fabio Q. B. da Silva and Abdelrahman Saher},
	booktitle = {Proceedings of the 12\textsuperscript{th} {ACM}/{IEEE} International Symposium on Empirical Software Engineering and Measurement - {ESEM} 18},
	title     = {Computer games are serious business and so is their quality},
	year      = {2018},
	doi       = {10.1145/3239235.3268923},
}

@InProceedings{delaurentisAutomatedGameBalance,
	author    = {DeLaurentis, Daniel A. and Panchal, Jitesh H. and Raz, Ali K. and Balasubramani, Prajwal and Maheshwari, Apoorv and Dachowicz, Adam and Mall, Kshitij},
	booktitle = {2021 IEEE Conference on Games (CoG)},
	title     = {Toward Automated Game Balance: A Systematic Engineering Design Approach},
	year      = {2021},
	doi       = {10.1109/CoG52621.2021.9619032},
}

@InProceedings{demesentiersilvaAIbasedPlaytestingContemporary2017,
	author    = {{de Mesentier Silva}, Fernando and Lee, Scott and Togelius, Julian and Nealen, Andy},
	booktitle = {Proceedings of the 12th {{International Conference}} on the {{Foundations}} of {{Digital Games}}},
	title     = {{{AI-based}} Playtesting of Contemporary Board Games},
	year      = {2017},
	address   = {{Hyannis Massachusetts}},
	publisher = {{ACM}},
	doi       = {10.1145/3102071.3102105},
}

@InProceedings{gudmundssonHumanLikePlaytestingDeep2018,
	author    = {Gudmundsson, Stefan Freyr and Eisen, Philipp and Poromaa, Erik and Nodet, Alex and Purmonen, Sami and Kozakowski, Bartlomiej and Meurling, Richard and Cao, Lele},
	booktitle = {2018 {{IEEE Conference}} on {{Computational Intelligence}} and {{Games}} ({{CIG}})},
	title     = {Human-{{Like Playtesting}} with {{Deep Learning}}},
	year      = {2018},
	publisher = {{IEEE}},
	doi       = {10.1109/CIG.2018.8490442},
}

@Article{isaksenExploringGameSpace2018,
	author   = {Isaksen, A. and Gopstein, D. and Togelius, J. and Nealen, A.},
	journal  = {IEEE Transactions on Games},
	title    = {Exploring {{Game Space}} of {{Minimal Action Games}} via {{Parameter Tuning}} and {{Survival Analysis}}},
	year     = {2018},
	doi      = {10.1109/TCIAIG.2017.2750181},
}

@InProceedings{pfauDungeonsReplicantsAutomated2020,
	author     = {Pfau, J. and Liapis, A. and Volkmar, G. and Yannakakis, G. N. and Malaka, R.},
	booktitle  = {2020 {{IEEE Conference}} on {{Games}} ({{CoG}})},
	title      = {Dungeons {{Replicants}}: {{Automated Game Balancing}} via {{Deep Player Behavior Modeling}}},
	year       = {2020},
	doi        = {10.1109/CoG47356.2020.9231958},
}

@InProceedings{roohiPredictingGameDifficulty2020,
	author    = {Roohi, Shaghayegh and Relas, Asko and Takatalo, Jari and Heiskanen, Henri and H{\"a}m{\"a}l{\"a}inen, Perttu},
	booktitle = {Proceedings of the Annual Symposium on Computer-Human Interaction in Play},
	title     = {Predicting game difficulty and churn without players},
	year      = {2020},
}

@InProceedings{silvaAIEvaluatorSearch,
	author    = {Fernando de Mesentier Silva and Scott Lee and Julian Togelius and Andy Nealen},
	booktitle = {AAAI Workshops},
	title     = {AI as Evaluator: Search Driven Playtesting of Modern Board Games},
	year      = {2017},
}

@InProceedings{zhengWujiAutomaticOnline2019,
	author     = {Zheng, Y. and Xie, X. and Su, T. and Ma, L. and Hao, J. and Meng, Z. and Liu, Y. and Shen, R. and Chen, Y. and Fan, C.},
	booktitle  = {2019 34th {{IEEE}}/{{ACM International Conference}} on {{Automated Software Engineering}} ({{ASE}})},
	title      = {Wuji: {{Automatic Online Combat Game Testing Using Evolutionary Deep Reinforcement Learning}}},
	year       = {2019},
	doi        = {10.1109/ASE.2019.00077},
}

@InProceedings{politowskiAreOldDays2016,
	author     = {Politowski, Cristiano and Fontoura, Lisandra and Petrillo, Fabio and Gu{\'e}h{\'e}neuc, Yann-Ga{\"e}l},
	booktitle  = {Proceedings of the 5th {{International Workshop}} on {{Games}} and {{Software Engineering}} - {{GAS}} '16},
	title      = {Are the Old Days Gone?: A Survey on Actual Software Engineering Processes in Video Game Industry},
	year       = {2016},
	address    = {{Austin, Texas}},
	publisher  = {{ACM Press}},
	doi        = {10.1145/2896958.2896960},
	langid     = {english},
	shorttitle = {Are the Old Days Gone?},
}

\end{document}